\documentclass[sigconf]{acmart}

\usepackage{graphicx}
\usepackage{multirow}
\usepackage{url}
\usepackage{amsmath}
\setcopyright{none}
\acmDOI{}

\acmISBN{}

\acmConference[]{Website:  http://bugtriage.mybluemix.net/}
\acmYear{}

\graphicspath{{./images/pdf/}}
\begin{document}
\title{DeepTriage: Exploring the Effectiveness of Deep Learning for Bug Triaging}
\author{
	Senthil Mani, Anush Sankaran, Rahul Aralikatte
}
\affiliation{IBM Research, India}
\email{{sentmani, anussank, rahul.a.r}@in.ibm.com}

% The default list of authors is too long for headers.
\renewcommand{\shortauthors}{Senthil Mani et al.}

\begin{abstract}
	For a given software bug report, identifying an appropriate developer who could potentially fix the bug is the primary task of a bug triaging process. A bug title (summary) and a detailed description is present in most of the bug tracking systems. Automatic bug triaging algorithm can be formulated as a classification problem, which takes the bug title and description as the input, mapping it to one of the available developers (class labels). The major challenge is that the bug description usually contains a combination of free unstructured text, code snippets, and stack trace making the input data highly noisy. The existing bag-of-words (BOW) feature models do not consider the syntactical and sequential word information available in the unstructured text.
	
	In this research, we propose a novel bug report representation algorithm using an attention based deep bidirectional recurrent neural network (DBRNN-A) model that learns a syntactic and semantic feature from long word sequences in an unsupervised manner. Instead of BOW features, the DBRNN-A based bug representation is then used for training the classifier. Using an attention mechanism enables the model to learn the context representation over a long word sequence, as in a bug report. To provide a large amount of data to learn the feature learning model, the unfixed bug reports (constitute about 70\% bugs in an open source bug tracking system) are leveraged upon as an important contribution of this research, which were completely ignored in the previous studies. 
	Another major contribution is to make this research reproducible by making the source code available and creating a public benchmark dataset of bug reports from three open source bug tracking system: Google Chromium, Mozilla Core, and Mozilla Firefox. For our experiments, we use 383,104 bug reports from Google Chromium, 314,388 bug reports from Mozilla Core, and 162,307 bug reports from Mozilla Firefox. Experimentally we compare our approach with BOW model and softmax classifier, support vector machine, naive Bayes, and cosine distance and observe that DBRNN-A provides a higher rank-10 average accuracy.
\end{abstract}

% The code below should be generated by the tool at
% http://dl.acm.org/ccs.cfm
% Please copy and paste the code instead of the example below. 

%\begin{CCSXML}
%	<ccs2012>
%	<concept>
%	<concept_id>10010147.10010257.10010321</concept_id>
%	<concept_desc>Computing methodologies~Machine learning algorithms</concept_desc>
%	<concept_significance>300</concept_significance>
%	</concept>
%	<concept>
%	<concept_id>10011007</concept_id>
%	<concept_desc>Software and its engineering</concept_desc>
%	<concept_significance>300</concept_significance>
%	</concept>
%	</ccs2012>
%\end{CCSXML}
%
%\ccsdesc[300]{Computing methodologies~Machine learning algorithms}
%\ccsdesc[300]{Software and its engineering~Software organization and properties}

 %End generated code

%\keywords{Deep Learning, bug triaging, recurrent neural network, open source code}

\maketitle

\begin{figure}[t]
	\centering
	\includegraphics[width=3.3in]{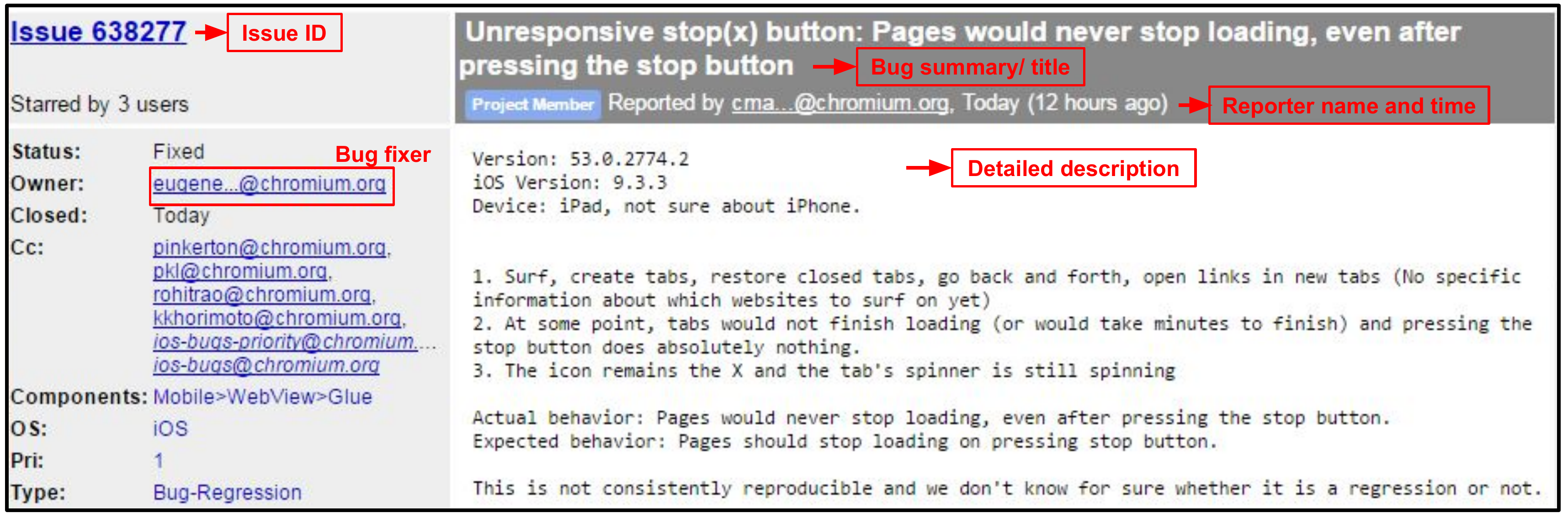}
	\caption{Screenshot of a bug report available in Google Chromium project, bug ID: 638277. The bug report usually consists of a brief summary and a detailed description at the time of reporting.}
	\label{fig:screenshot}
\end{figure}
\section{Introduction}
In an usual process, the end user encounters a bug (also called an issue or a defect) while working on the system and reports the issue in a bug tracking system~\cite{fischer2003populating}. Fig~\ref{fig:screenshot} shows a sample screenshot of a bug reported in Google Chromium project (bug ID: 638277). The bug report usually contains a bug summary and a detailed description mentioning the steps to reproduce. Bugs with \textit{fixed} status also contains the developer who fixed the bug and is called as the owner.  
The process of bug triaging consists of multiple steps where first step primarily involves assigning the bug to one of the developers who could potentially solve the bug. Thus, in the rest of this research bug triaging refers to the task of developer assignment for the bug report~\cite{anvik2006should}. In large scale systems, with a huge amount of incoming bugs, manually analyzing and triaging a bug report is a laborious process. Manual bug triaging is usually performed using the bug report content, primarily consisting of the summary and description. While additional sources of input has been explored in the literature such as developer profiling from github~\cite{badashian2015crowdsourced} and using component information~\cite{bhattacharya2010fine}, majority of the research efforts have focused on leveraging the bug report content for triaging~\cite{anvik2011reducing}~\cite{Jonsson2016} \cite{shokripour2013so}~\cite{tamrawi2011fuzzy}  \cite{wang2014fixercache} \cite{xuan2015towards} \cite{xuan2012developer}. Using the bug report content, automated bug triaging can be formulated as a classification problem, mapping the bug title and description to one of the developers (class labels). However, the bug report content contains noisy text information including code snippets, and stack trace details, as observed in Fig.~\ref{fig:screenshot}. Processing such unstructured and noisy text data is a major challenge in learning a classifier. 

\begin{figure}[ht]
	\centering
	\includegraphics[width=3.3in]{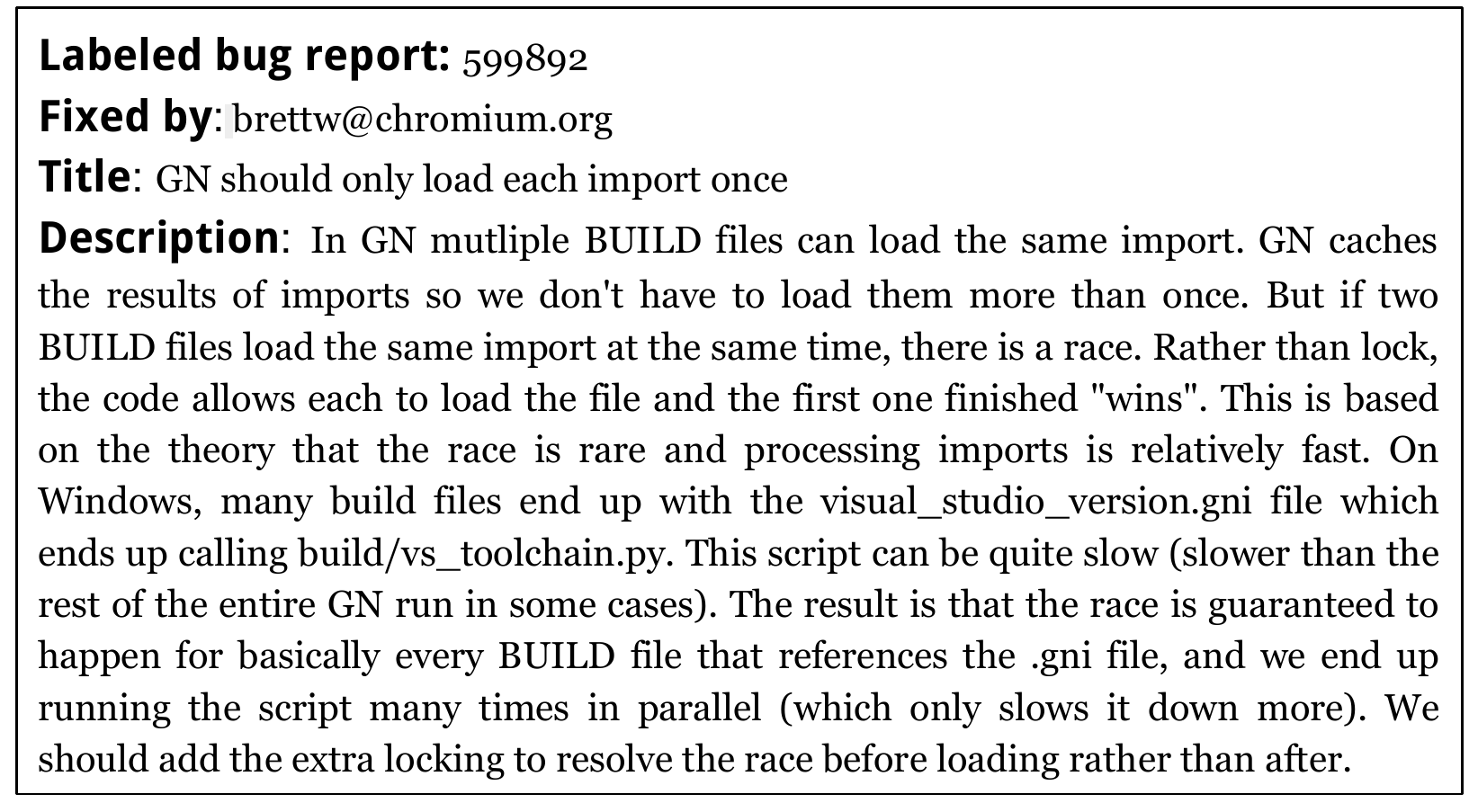}
	\caption{A bug report from Google Chromium bug repository used as a labeled template for training the classifier.}
	\label{fig:trainbug}
\end{figure}

\subsection{Motivating Example}
\begin{figure}[t]
	\centering
	\includegraphics[width=3.3in]{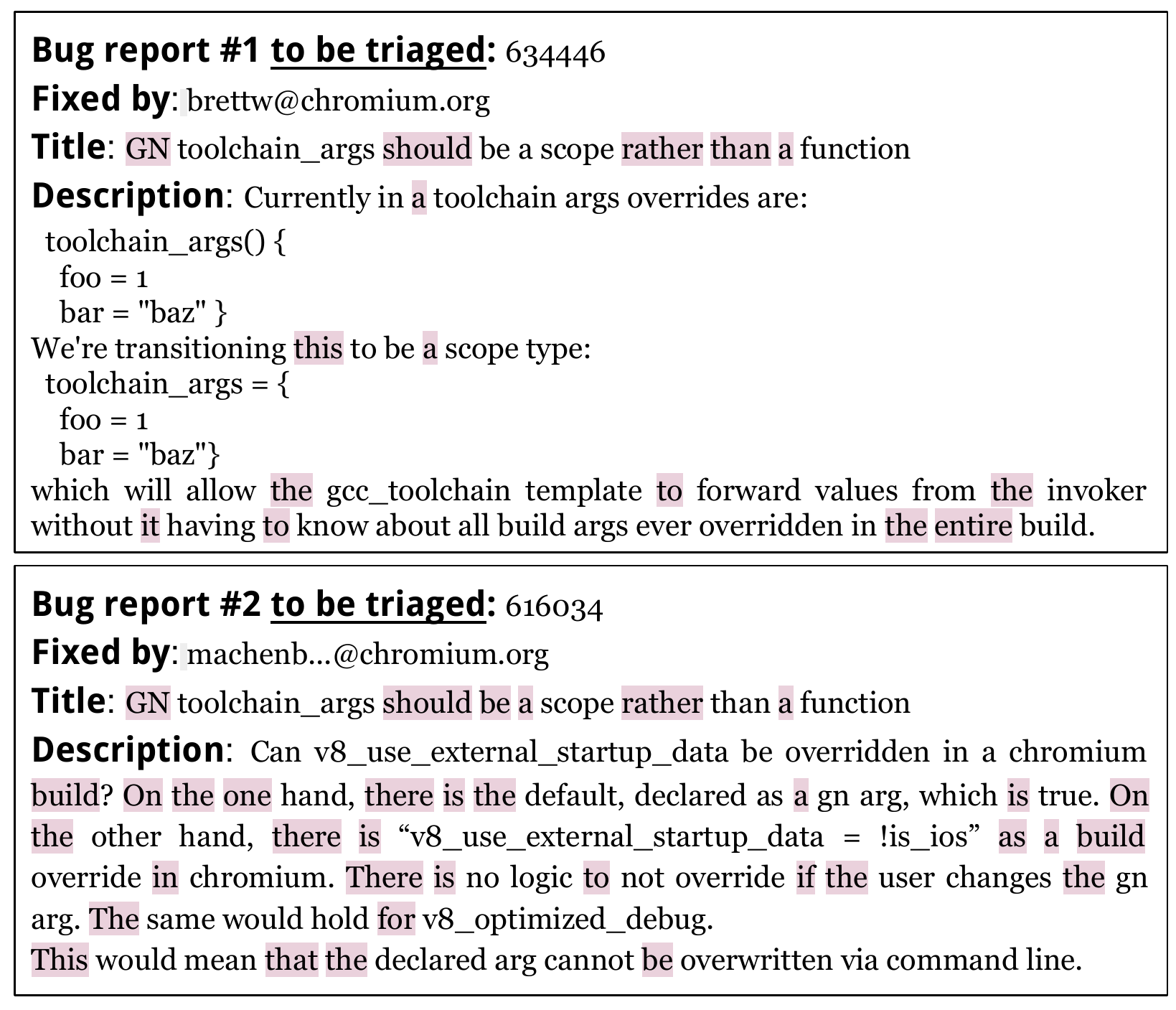}
	\caption{Two example bug reports from Google Chromium bug repository for which a suitable developer has to be predicted. By ground truth, bug report \#1 was fixed by the same developer as in the training instance. However, the BOW feature of bug report \#2 is more similar to the training instance than the bug report \#1. The overlapping words with the training bug are highlighted.}
	\label{fig:testbug}
\end{figure}

Consider a labeled bug report example shown in Fig.~\ref{fig:trainbug}. The bag-of-words (BOW) feature representation of the bug report creates a boolean array marking true (or term-frequency) for each vocabulary word in the bug report~\cite{anvik2011reducing}. During training, a classifier will learn this representation to the corresponding class label \textit{brettw@chromium.org}. For the given two testing examples shown in Fig.~\ref{fig:testbug}, the actual fixer of first example, with bug id $634446$, is \textit{brettw@chromium.org} while the second example bug with id $616034$ was fixed by \textit{machenb...@chromium.org}. However, based on BOW features there are 12 words common between testing report\#1 and the train report, while there are 21 words common between testing report\#2 and the train report. Hence, a BOW model mis-classifies that the testing bug report\#2 with id $616034$ should be fixed by \textit{brettw@chromium.org}. 
The reasons for the mis-classification are: (i) BOW feature model considers the sentence as a bag-of-words loosing the ordering (context) of words, and (ii) the semantic similarity between synonymous words in the sentence are not considered.
Even though a bag-of-n-grams model considers a small context of word ordering, they suffer from high dimensionality and sparse data~\cite{hindle2012naturalness}.
The semantic similarity between word tokens can be learnt using a skip-gram based neural network model called 
\textit{word2vec}~\cite{mikolov2013efficient}. This model relies on distributional hypothesis which claims that words that appear in the same context in the sentence share a semantic meaning. Ye et al.,~\cite{Ye2016} built a shared word representation using \textit{word2vec} for word tokens present in code language and word tokens present in descriptive language. The main disadvantage of \textit{word2vec} is that it learns a semantic representation of individual word tokens, however, does not consider a sequence of word tokens such as a sentence. An extension of \textit{word2vec} called \textit{paragraph vector}~\cite{le2014distributed} considers the ordering of words, but only for a small context.  

The rest of the paper is organized as follows: section 2 highlights the main research questions addressed in this research work and the key contributions, section 3 details the proposed approach including the deep learning algorithm and the classifier, section 4 talks about the experimental data collected in this research, section 5 discuss our experimental results and analysis, section 6 discusses some of the threats to validate our claims, section 7 talks about other applications that can be addressed using the proposed feature learning algorithm, section 8 explains about some closely related work and section 9 concludes our work with some future directions.

\begin{figure*}[t]
	\centering
	\includegraphics[width=5.5in]{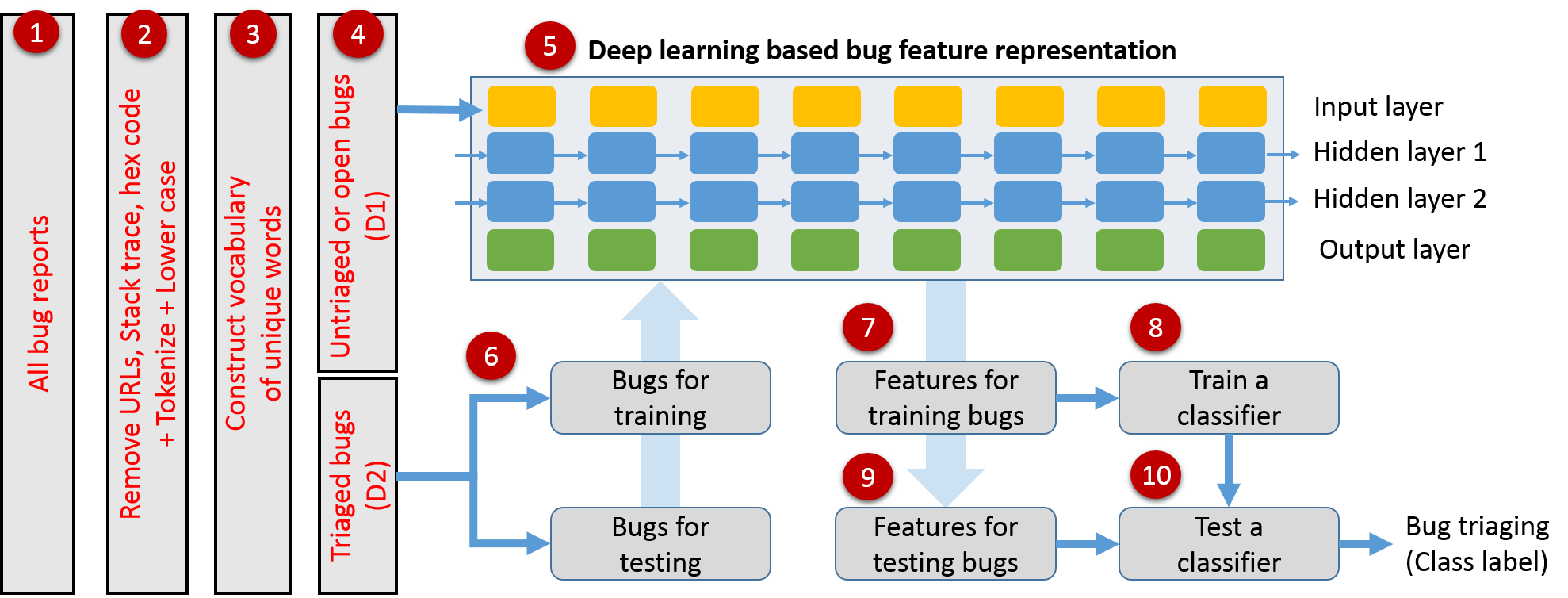}
	\caption{The flow diagram of the overall proposed algorithm highlighting the important steps.}
	\label{fig:overall}
\end{figure*}

\section{Research Contributions}

Learning semantic representation from large pieces of text (such as in description of bug reports), preserving the order of words, is a challenging research problem. Thus, we propose a deep learning technique, which will learn a succinct fixed-length representation of a bug report content in an unsupervised fashion ie., the representation will be learnt directly using the data without the need for manual feature engineering. The main research questions (RQ) that we are trying address in this research paper are as follows:
\begin{enumerate}
	\item \textbf{RQ1}: Is it feasible to perform automated bug triaging using deep learning?
	\item \textbf{RQ2}: How does the unsupervised feature engineering approach perform, compared to traditional feature engineering approaches? 
	\item \textbf{RQ3}: Is there an effect on the number of training samples per class on the performance of the classifier?
	\item \textbf{RQ4}: What is the effect of using only the title (or summary) of the bug report in performing triaging compared with the using the description as well ?
	\item \textbf{RQ5}: Is transfer learning effective using deep learning, where the deep learning model is trained using one dataset and used to perform triaging in another dataset?
	 
\end{enumerate}

Recently, recurrent neural network (RNN) based deep learning algorithms have revolutionized the concept of word sequence representation and have shown promising breakthroughs in many applications such as language modeling and machine translation. Lam et al.~\cite{7372035} used deep neural network (DNN) with rSVM to learn a common representation between source code and the bug reports and used it for effective bug localization. White et al.,~\cite{7180092} provided a broad perspective on how deep learning can be used in software repositories to solve some challenging problems. The main contributions of this research are summarized as follows:
\begin{itemize}
	\item A novel bug report representation approach is proposed using DBRNN-A: Deep Bidirectional Recurrent Neural Network with Attention mechanism and with Long Short-Term Memory units (LSTM)~\cite{pham2014dropout}. The proposed deep algorithm algorithm is capable of remembering the context over a long sequence of words.
	\item The untriaged and unsolved bug reports constitute about 70\% in an open source bug repository and are usually ignored in the literature~\cite{Jonsson2016}. In this research, we provide a mechanism to leverage all the untriaged bugs to learn bug representation model in an unsupervised manner.
	\item Experimental data (bug reports) are collected from three open source bug repositories: $3,83,104$ from     Chromium, $3,14,388$ from Mozilla Core, and $1,62,307$ from Mozilla Firefox. Performance of the classifiers trained on different train-test splits of datasets~\cite{Jonsson2016}~\cite{LamkanfiMSR13} are neither comparable nor reproducible. Thus, to enable our research reproducible, the entire dataset along with the exact train test split and the source code of our approach are made publicly available for research purpose\footnote{Made available at: http://bugtriage.mybluemix.net/}.
	\item We further study the effectiveness of the proposed bug training in a cross-data testing scenario (transfer learning). By training the model with bugs from Chromium project and re-using the model for triaging bugs in Core and Firefox projects (Mozilla bug repository), the transfer learning ability of the deep learning model is showcased.
\end{itemize}

\begin{figure*}[ht]
	\centering
	\includegraphics[width=6.7in]{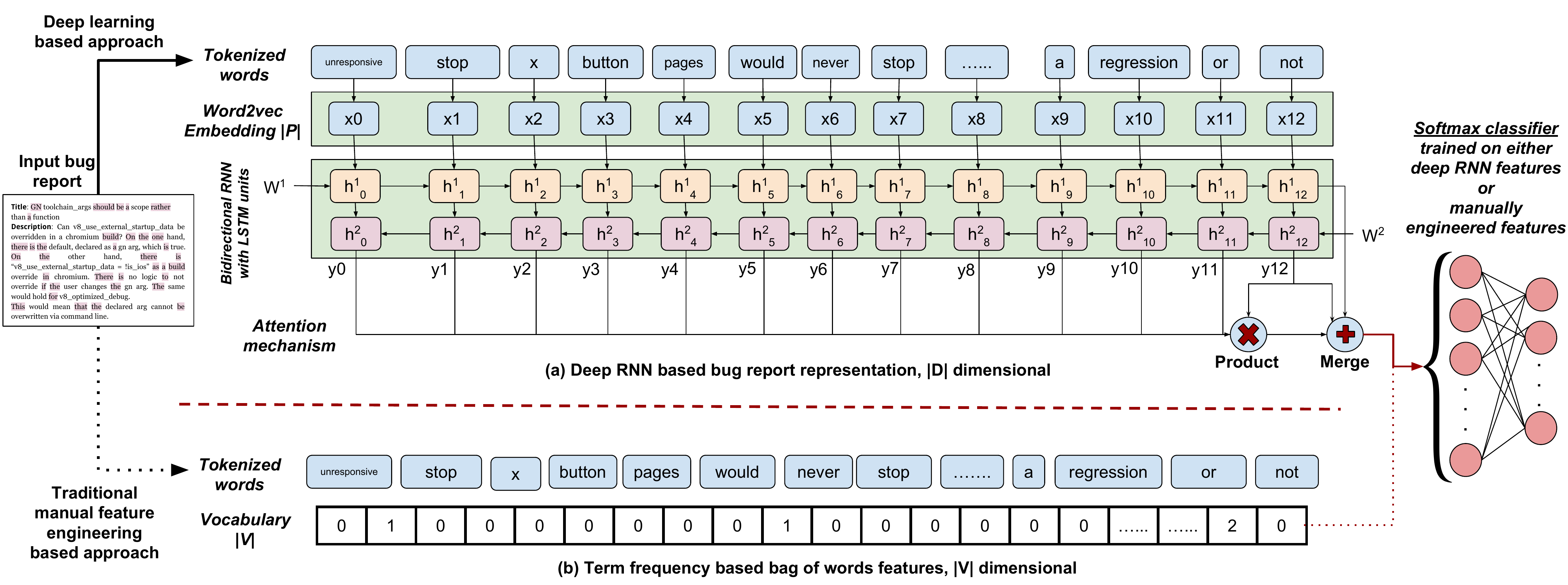}
	\caption{Detailed explanation of the working of a deep bidirectional Recurrent Neural Network (RNN) with LSTM units for an example bug report shown in Fig.~\ref{fig:screenshot}. It can be seen that the deep network has multiple hidden layers, learning a complex hierarchical representation from the input data. As a comparison, tf based bag-of-words (BOW) representation for the same example sentence is also shown.}
	\label{fig:rnn}
\end{figure*}

\section{Proposed Approach}
The problem of automated bug triaging of software bug reports is formulated as a supervised classification approach with the input data being the bug summary (title) and the bug description. Fig.~\ref{fig:overall} highlights the major steps involved the proposed automated bug triaging algorithm and are explained as follows: 
\begin{enumerate}
	\item a bug corpus having title, description, reported time, status, and owner is extracted from an open source bug tracking system,
	\item handling the URLs, stack trace, hex code, and the code snippets in the unstructured description requires specialized training of the deep learning model, and hence in this research work, those contents are removed in the preprocessing stage,
	\item a set of unique words that occurred for at least $k$-times in the corpus is extracted as the vocabulary,
	\item the triaged bugs (D2) are used for classifier training and test, while all the untriaged/ open bugs (D1) are used to learn a deep learning model,
	\item a deep bidirectional recurrent neural network with attention mechanism technique learns a bug representation considering the combined bug title and description as a sequence of word tokens,
	\item the triaged bugs (D2) are split into train and test data with a $10$ fold cross validation to remove training bias,
	\item feature representation for the training bug reports are extracted using the learnt DB-RNN algorithm,
	\item a supervised classifier is trained for performing developer assignment as a part of bug triaging process,
	\item feature representation of the testing bugs are then extracted using the learnt deep learning algorithm,
	\item using the extracted features and the learnt classifier, a probability score for every potential developer is predicted and the accuracy is computed in the test set.
\end{enumerate}
The proposed approach varies with the traditional pipeline for automated bug triaging in (i) step 4 where the untriaged bugs (D1) are completely ignored and (ii) the deep learning based bug report representation instead of a bag-of-words representation. Addition of steps 4 and 5, enables to automatically learn a bug report representation from the data itself instead of manual engineering.

\subsection{Deep Bidirectional Recurrent Neural Network with Attention (DBRNN-A)}
The proposed DBRNN-A based feature learning approach, as shown in Fig.~\ref{fig:rnn}, have the following key advantages:
\begin{itemize}
	\item DBRNN-A can learn sentence representation preserving the order and syntax of words, as well as, retaining the semantic relationship. Long short-term memory (LSTM) cells~\cite{hochreiter1997long} are used in the hidden layer which have a memory unit that can remember longer word sequences and also can solve the vanishing gradient problem~\cite{sak2014long}.
	\item Intuitively, all the words in the content may not be useful in triaging the bug. To incorporate this, we introduce an attention mechanism~\cite{luong2015effective} which learns to ``attend" to only the important words in the bug report during classification. As the attention mechanism chooses only a few words during classification, the DBRNN-A can learn context representation from really long word sequences.
	\item A bidirectional RNN~\cite{graves2013speech} considers the word sequence both in forward direction (first word to last word) and in backward direction (last word to first word) and merges both these representations. Thus, a context of a particular word includes both the previous few words and following few words making the representation more robust.
\end{itemize}

For each word, a one-hot $|V|$-dimensional representation is extracted using the vocabulary \footnote{http://www.wildml.com/2015/09/recurrent-neural-networks-tutorial-part-1-introduction-to-rnns/}, over which a a $|P|$-dimensional \textit{word2vec} representation~\cite{mikolov2013distributed} is learnt. As shown in Fig.~\ref{fig:rnn} (a), a DBRNN-A with LSTM units is learnt over this word representation, to obtain a $|D|$-dimensional feature representation of the entire bug report (title + description). RNN is a sequence network containing a hidden layer with $m$ hidden units, $\mathbf{h} = \{\mathbf{h_1}, \mathbf{h_2}, \ldots, \mathbf{h_m}\}$. The input to the system is a sequence of word representations, $\mathbf{x} = \{\mathbf{x_1}, \mathbf{x_2}, \ldots, \mathbf{x_m}\}$, and produces a sequence of outputs $\mathbf{y} = \{\mathbf{y_1}, \mathbf{y_2}, \ldots, \mathbf{y_m}\}$. 
Each hidden unit is a state model converting the previous state, $s_{i-1}$ and a word, $x_i$ to the next state, $s_i$ and an output word, $y_i$. The term ``recurrent" explains that every hidden unit performs the same function in recurrent fashion, $f: \{s_{i-1}, x_i\}\rightarrow \{s_i,y_i\}$. Intuitively, the state $s_i$ carries the cumulative information of the $i$ previous words observed. The output $y_m$ obtained from the last hidden node is a cumulative representation of the entire sentence. For example, consider the tokenized input sentence provided in Fig.~\ref{fig:rnn}. When $i=1$, $x_i$ is the $|P|$-dimensional \textit{word2vec} representation of the input word, \emph{unresponsive} and the previous state $s_0$ is randomly initialized. Using the LSTM function $f$, the current state $s_1$ and the word output $y_1$ are predicted. Given the next word \emph{stop} and the current state $s_1$, the same function $f$ is used to predict $s_2$ and $y_2$. The shared function reduces the number of learnable parameters as well as retains the context from all the words in the sequence. For language modeling or learning sentence representation, the ground truth $y_i$ are the next word in the sequence $x_{i+1}$, that is, upon memorizing the previous words in the sentence the network tries to predict the next word. LSTM function~\cite{graves2013speech} have special purpose built-in memory units to store the context information over longer sentences. 

Further, to selectively remember and learn from the important words in a bug report, an attention model is employed. An attention vector is derived by performing a weighted summation of all the computed outputs, $y_i$, as follows:
\begin{equation}
a_m = \sum_{i=1}^{m}\alpha_i y_i
\end{equation}
Intuitively, $\alpha_i$ associates a weight to each word implying the importance of that word for classification.
Two different deep RNN based feature model are learnt, one with input word sequence running forward and one with input word sequence running backward. The final representation, $r$, obtained for a bug report, is provided as follows:
\begin{equation}
r = \underbrace{y_m \oplus a_m}_\text{\text{forward LSTM}} \oplus \underbrace{y_m \oplus a_m}_\text{\text{backward LSTM}}
\end{equation} 
where $\oplus$ represents concatenation of the vectors. In comparison as shown in Fig.~\ref{fig:rnn} (b), a term frequency based BOW model would produce a $|V|$-dimensional representation for the same bug report, where $V$ is the size of vocabulary. Typically, the size of $|P|$ is chosen as $300$~\cite{mikolov2013distributed} and the size of $D$ will be less than $4|P|$ ($<1200$) is much smaller than the size of $|V|$. For example, consider $10,000$ bugs used for training with $250,000$ unique words ($|V|$). BOW model representation would produce a sparse feature matrix of size $10,000 \times 250,000$, while the proposed DBRNN-A would produce a dense and compact representation with a feature matrix of size $10,000 \times 1,200$.

The entire deep learning model was implemented in Python using Keras library. To the best of our knowledge, this is the first time a deep sequence learning model has been applied to learn a bug representation and used those features to learn a supervised model for automated software bug triaging.

\subsection{Classifying (Triaging) a Bug Report}
The aim of the supervised classifier is to learn a function, $C$, that maps a bug feature representation to a set of appropriate developers. Formulating automated bug triaging as a supervised classification problem has been well established in literature~\cite{bhattacharya2010fine} \cite{xuan2015towards}. However, it is well understood that a classification is only as good as the quality of features. Hence, the major contribution in this research is to propose a better bug report representation model and to improve the performance of existing classifiers. In this research we use a softmax classifier, a popular choice of classifier along with deep learning~\cite{graves2013speech}~\cite{cho2014learning}~\cite{graves2005framewise}. Softmax classifier is a generalization of logistic regression for multiclass classification, taking the features and providing a vector of scores with length equal to the number of the classes. A softmax classifier normalizes these score values and provides an interpretable probability value of the $i$-th bug report belonging to the class.

\section{Large Scale Public Bug Triage Dataset}
A huge corpus of bug report data is obtained from three popular open source system : Chromium\footnote{https://bugs.chromium.org/p/chromium/issues/list}, Mozilla Core, and Mozilla Firefox\footnote{https://bugzilla.mozilla.org/} and the data collection process is explained in this section. To make this research reproducible, the entire data along with the exact train-test protocol and with source code is made available at: \url{http://bugtriage.mybluemix.net/}.

\begin{table}[!t]
	\centering
	\small
	\begin{tabular}{|l|c|c|c|}
		\hline
		\textbf{Property} & \textbf{Chromium} & \textbf{Core}  & \textbf{Firefox}\\ \hline
		Total bugs & 383,104  & 314,388  & 162,307   \\ \hline
		Bugs for learning feature &  263,936 & 186,173  &  138,093  \\ \hline
		Bugs for classifier &  118,643 & 128,215  &  24,214  \\ \hline
		Vocabulary size $|V|$ & 71,575  & 122,578  &  57,922  \\ \hline
	\end{tabular}
	\caption{\label{tab:deep0} Summary of the three different bug repositories, Google Chromium, Mozilla Core, and Mozilla Firefox used in our experiments.}
\end{table}

\subsection{Data Extraction}
Bug reports from the Google Chromium project were downloaded for the duration of August 2008 (Bug ID: 2) - July 2016 (Bug ID: 633012). A total of 383,104 bugs where collected with the bug title, description, the bug owner, and the reported time. The developer in the ``owner" field is considered as the ground truth triage class for the given bug\footnote{\url{https://www.chromium.org/for-testers/bug-reporting-guidelines/triage-best-practices}}. Bugs with status as \emph{Verified} or \emph{Fixed}, and type as \emph{bug}, and has a valid ground truth bug owner are used for training and testing the classifier while rest of the bugs are used for learning a bug representation. However, we noticed that there were a total of 11,044 \emph{bug} reports with status as \emph{Verified} or \emph{Fixed} and did not have a valid owner associated. These bugs are considered as open bugs, resulting in a total of 263,936 ($68.9\%$) bug reports are used for deep learning, and 118,643 ($31\%$) bugs are used for the classifier training and testing.

Data from two popular components from Mozilla bug repository are extracted: Core and Firefox. 314,388 bug reports are extracted from Mozilla Core reported between April 1998 (Bug ID: 91) and June 2016 (Bug ID: 1278040), and 162,307 bug reports are extracted from Mozilla Firefox reported between July 1999 (Bug ID: 10954) and June 2016 (Bug ID: 1278030). The developer in the ``Assigned To" %\footnote{\url{https://wiki.mozilla.org/BMO/UserGuide/BugFields#assigned_to}}
 is considered as the ground truth triage class during classification. Bug reports with status as \emph{verified fixed}, \emph{resolved fixed}, and \emph{closed fixed} are used for classifier training and testing. However, some of the \emph{fixed} reports did not have a developer assigned to it, such as, in Core ($7219/135434=5.33\%$) and in Firefox ($3716/27930=13.3\%$). After ignoring these bugs, a final number of $1,28,215$ bugs for Core and $24,214$ bugs for Firefox are considered for classifier training and testing. The summary of the datasets is provided in Table~\ref{tab:deep0}.

\subsection{Data Preprocessing}
The three datasets are preprocessed independently using the same set of steps and a benchmark protocol is created. For every bug report, the title and description text content of the bug are combined. Preprocessing of the unstructured textual content involves removing URLs, hex code, and stack trace information, and converting all text to lower case letters. Tokenization of words is performed using Stanford's \textit{NLTK} package\footnote{http://www.nltk.org/api/nltk.tokenize.html}.A vocabulary of all words is constructed using the entire corpus. 
To remove rarely occurring words and reduce the vocabulary size, usually the top-$F$ frequent words are considered or only those words occurring with a minimum frequency are considered~\cite{Ye2016}. For the extracted data, we experimentally observed that a minimum word frequency of $5$ provided a good trade-off between the vocabulary size and performance.

\subsection{Training Data for Deep Learning}
In our data split mechanism, the classifier testing data is unseen data and hence cannot be used for the deep learning algorithm. A  design choice was taken for not using the classifier training data for training the deep learning model, as including them only marginally improved the accuracy but largely increased the training time. Thus, only the untriaged bugs (explained in the data extraction subsection) is used for training the deep learning model. Also, using a non-overlapping dataset for training the feature model and training the classifier model highlights the generalization ability of the features.

\subsection{Training Data for Classification}
For training and testing the supervised classifier, a 10-fold cross validation model as proposed by Betternburg et al~\cite{bettenburg2008duplicate} is followed. All the fixed bug reports are arranged in chronological order and split into $11$ sets. Starting from the second fold, every fold is used as a test set, with the cumulation of previous folds for training. Typically, in an open source project the developers keep changing overtime, and hence chronological splitting ensures that the train and test sets have highly overlapping developers.
Further, in order to make the training effective, we need more number of training sample per developer. In a recent study, Jonsson et al.,~\cite{Jonsson2016} trained using those developers who have at least addressed $50$ bug reports i.e., minimum number of training samples per class is $50$. From different studies in the literature~\cite{anvik2011reducing}~\cite{bhattacharya2010fine}, it is clear that the threshold parameter affect the classification performance. Thus, in this research we study the direct relation between the threshold value and the classification performance, by having four different thresholds for the minimum number of training samples per class as $0$, $5$, $10$, $20$. To perform a closed training experiment, it is made sure that all the classes available in testing are available for training while there are additional classes in training which are not available in the test set. Thus, for every test bug report with an owner, the classifier is already trained with other bugs trained by the same owner.

\section{Experimental Evaluation}
\begin{figure}[t]
	\centering
	\includegraphics[width=3.3in]{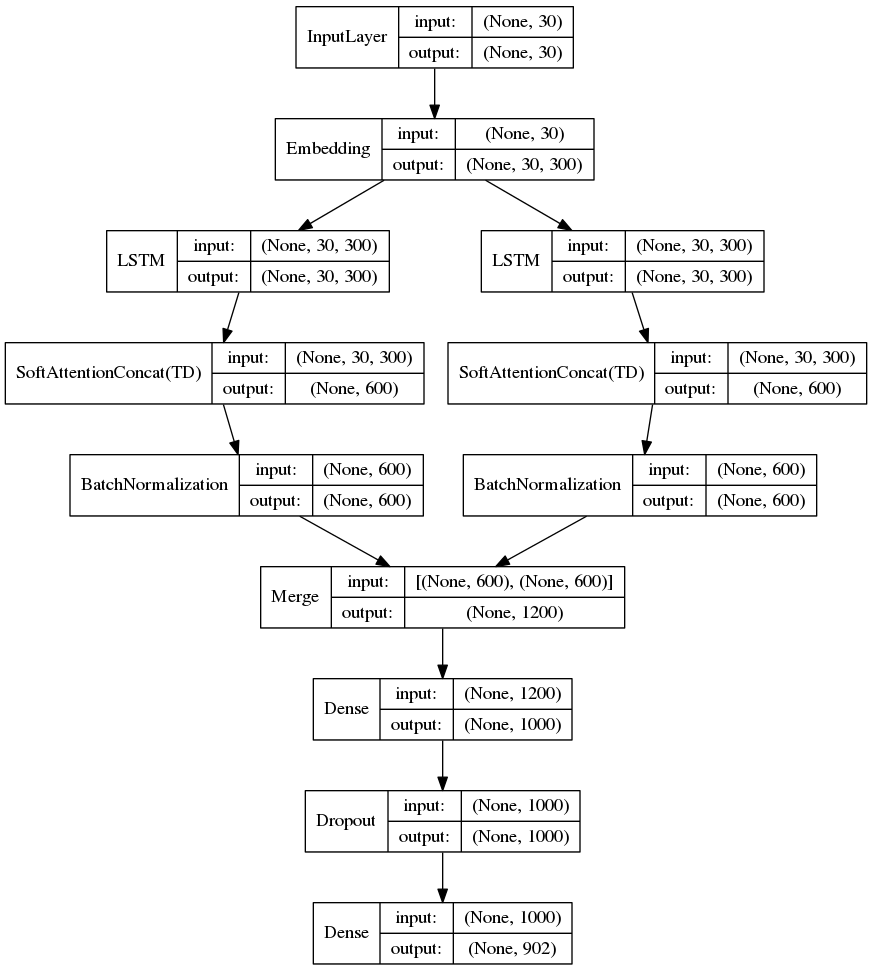}
	\caption{The architecture of DBRNN-A detailing all the parameters of the model.}
	\label{fig:dbrnna}
\end{figure}

\begin{table*}[ht]
	\centering
	\begin{tabular}{|p{1.4cm}|l|c|c|c|c|c|c|c|c|c|c|c|}
		\hline
		\textbf{Threshold}  & \textbf{Classifier}  & \textbf{CV\#1}  & \textbf{CV\#2} & \textbf{CV\#3} & \textbf{CV\#4} & \textbf{CV\#5} & \textbf{CV\#6} & \textbf{CV\#7} & \textbf{CV\#8} & \textbf{CV\#9} & \textbf{CV\#10} & \textbf{Average}\\ \hline
		\multirow{5}{*}{\parbox{1.4cm}{Min. train samples per class = 0}} & BOW + MNB &  21.9 & 25.0  & 26.0  & 23.0  &  23.7 & 25.9  & 26.3  & 26.1  & 28.7  & 33.3  & 26.0 $\pm$ 3.0   \\ \cline{2-13}
		& BOW + Cosine &  18.4  & 20.1  & 21.3  & 17.8  & 20.0  & 20.6  & 20.4  & 20.9  & 21.1  & 21.5  & 20.2 $\pm$ 1.2  \\ \cline{2-13}
		& BOW + SVM & 11.2   & 09.3  & 09.5  & 09.5  & 09.4  & 10.1  &  10.4 &  09.9 & 10.5  & 10.8  &  10.1 $\pm$ 0.6 \\ \cline{2-13}
		& BOW + Softmax & 12.5  & 08.5   & 08.6  & 08.7  &  08.6 &  08.5 &  09.1 &  08.9 &  08.7 &  08.7 & 09.1 $\pm$ 1.1  \\ \cline{2-13}
		& DBRNN-A + Softmax & 34.9  &  36.0 & 39.6  &  35.1 & 36.2  & 39.5  & 39.2  &  39.1 &  39.4 & 39.7  & \textbf{37.9 $\pm$ 1.9}\\ \hline \hline

		\multirow{5}{*}{\parbox{1.4cm}{Min. train samples per class = 5}} &BOW + MNB &  22.2  & 25.2  & 26.1  & 23.1  & 23.8  & 26.0  & 26.5  & 26.3  & 29.2  & 33.6  &  26.2 $\pm$ 3.1 \\ \cline{2-13}
		& BOW + Cosine &  18.6  & 20.2  & 21.4  & 18.2  &  19.1 & 20.7  & 21.1  & 21.0  &  21.6 & 22.0  &  20.4 $\pm$ 1.3 \\ \cline{2-13}
		& BOW + SVM &  11.3  &  11.1 & 08.1  & 08.3  &  09.2 & 09.0  & 08.9  & 08.7  &  08.5 &  09.0 & 09.2 $\pm$ 1.0  \\ \cline{2-13}
		& BOW + Softmax &  12.8  & 11.1  & 11.1  &  09.3 &  11.1 & 09.8  & 10.4  & 10.5  & 10.9  & 11.4  & 10.8 $\pm$ 0.9  \\ \cline{2-13}
		& DBRNN-A + Softmax &  32.2 & 33.2  & 37.0  &  36.4  &  37.1  & 37.2  &  38.3 & 39.0  & 39.1  & 38.2  & \textbf{36.8  $\pm$ 2.2} \\ \hline \hline

		\multirow{5}{*}{\parbox{1.4cm}{Min. train samples per class = 10}} &BOW + MNB &  22.4  &  25.5 &  26.4 & 23.3  & 24.1  & 26.5  & 26.8  & 27.0  & 30.1  &  34.3 & 26.6 $\pm$ 3.3  \\ \cline{2-13}
		
		& BOW + Cosine &  18.8 &  20.5  & 21.7  & 18.5  &  19.6 &  21.2 & 21.4  & 21.1 & 21.8  & 21.0  & 20.6 $\pm$ 1.3  \\ \cline{2-13}
		& BOW + SVM & 12.2  &  11.4  &  11.8 &  11.6 &  11.5 & 11.3  & 11.8  & 11.0  & 12.1  & 11.9  & 11.7 $\pm$ 0.4  \\ \cline{2-13}	
		& BOW + Softmax &  11.9  & 11.3  & 11.2   & 11.2  & 11.3  & 11.1  & 11.4  & 11.3  & 11.2  &  11.5 &  11.3 $\pm$ 0.2 \\ \cline{2-13}
		& DBRNN-A + Softmax &  36.2  & 37.1  & 40.45  & 42.2  & 41.2  & 41.3  &  44.0 &  44.3 &  45.3 & 46.0  & \textbf{41.8 $\pm$ 3.1}  \\ \hline \hline

		\multirow{5}{*}{\parbox{1.4cm}{Min. train samples per class = 20}} & BOW + MNB & 22.9  & 26.2  & 27.2  & 24.2  & 24.6  &  27.6 & 28.2  & 28.9  & 31.8  & 36.0  & 27.8 $\pm$ 3.7  \\ \cline{2-13}
		
		& BOW + Cosine &  19.3 &  20.9  & 22.2  & 19.4  & 20.0  &  22.3 & 22.3  & 22.9  & 23.1  & 23.0  & 21.5 $\pm$ 1.4  \\ \cline{2-13}
		& BOW + SVM &  12.2 & 12.0  & 11.9  & 11.9  &  11.6 & 11.5  & 11.3  & 11.6  & 11.6  &  11.9 & 11.7 $\pm$ 0.3  \\ \cline{2-13}	
		& BOW + Softmax & 11.9  & 11.8   & 11.4  & 11.3  &  11.2 & 11.1  & 11.0  & 11.8  &  11.3 & 11.7  & 11.5 $\pm$ 0.3  \\ \cline{2-13}
		& DBRNN-A + Softmax & 36.7  & 37.4 & 41.1  & 42.5  & 41.8  & 42.6  & 44.7  & 46.8  & 46.5  &  47.0 &  \textbf{42.7 $\pm$ 3.5} \\ \hline 
		
	\end{tabular}
	\caption{\label{tab:deep3} Rank-$10$ accuracy obtained on the Google Chromium project across the ten cross validations. The average accuracy over the cross validation and standard deviation is also reported. The best performing results are shown in bold.}
\end{table*}	

\begin{table*}[!h]
	\centering
	\begin{tabular}{|p{1.4cm}|l|c|c|c|c|c|c|c|c|c|c|c|}
		\hline
		\textbf{Threshold} & \textbf{Classifier}  & \textbf{CV\#1}  & \textbf{CV\#2} & \textbf{CV\#3} & \textbf{CV\#4} & \textbf{CV\#5} & \textbf{CV\#6} & \textbf{CV\#7} & \textbf{CV\#8} & \textbf{CV\#9} & \textbf{CV\#10} & \textbf{Average}\\ \hline
		\multirow{5}{*}{\parbox{1.4cm}{Min. train samples per class = 0}} & BOW + MNB &  21.6 &  23.6  & 29.7  &  30.3 &  31.0 &  31.2 &  31.9 & 31.7  &  32.3 &  32.1 & 29.5 $\pm$ 3.6  \\ \cline{2-13}
		
		& BOW + Cosine & 16.3  & 17.4   &  19.5 & 21.3  & 22.5  &  23.2 &  24.0 &  25.5 &  27.5 &  29.1 & 22.6  $\pm$ 3.9 \\ \cline{2-13}
		& BOW + SVM & 13.6  & 14.6   &  14.9 &  14.0 &  12.1 &  12.9 &  11.7 & 13.7  & 14.4  &  14.1 &  13.6 $\pm$  1.0\\ \cline{2-13}	
		&BOW + Softmax &  14.3 &   11.8 & 9.5  & 10.0  & 9.2  & 10.4  & 10.5  & 10.6  & 11.0  &  10.8 & 10.8 $\pm$ 1.4  \\ \cline{2-13}	
		& DBRNN-A + Softmax &  30.1  &  31.7 &  35.2 & 33.0  & 34.1  &  35.9 &  34.8 & 34.2  & 34.6  & 35.1  &  \textbf{33.9 $\pm$ 1.7} \\ \hline \hline
		
		\multirow{5}{*}{\parbox{1.4cm}{Min. train samples per class = 5}} & BOW + MNB & 20.7  &  23.8  & 29.7  &  31.4 &  31.7 &  33.8 &  35.6 &  36.7 & 35.8  & 36.2  & 31.5 $\pm$ 5.2  \\ \cline{2-13}
		
		& BOW + Cosine & 15.7  &  17.7  & 19.9  & 21.4  &  22.8 &  24.7 &  26.4 &  27.5 & 29.4  & 29.9  & 23.5 $\pm$ 4.6  \\ \cline{2-13}
		& BOW + SVM &  16.4 &  12.9  &  11.5 &  10.4 &  13.4 & 13.8  & 12.7  & 12.0  & 12.8  & 13.1  &  12.9 $\pm$ 1.5 \\ \cline{2-13}	
		&BOW + Softmax &  14.9 &   13.5 &  12.5   & 10.6  & 11.4  & 12.8  & 12.1  & 13.3  &  12.4 &  14.0 &  12.7 $\pm$ 1.2 \\ \cline{2-13}	
		& DBRNN-A + Softmax &  33.8  &  31.5 &  35.8 & 35.3  & 34.7  & 36.8  & 37.1  & 38.4  & 37.7  & 38.0  & \textbf{35.9 $\pm$ 2.1}  \\ \hline \hline

		\multirow{5}{*}{\parbox{1.4cm}{Min. train samples per class = 10}} & BOW + MNB & 18.4  &  23.9  & 29.8  &  33.4 & 36.7 & 39.4  &  38.5 &  40.8 & 41.3  & 42.5  &  34.5 $\pm$ 7.7 \\ \cline{2-13}
		
		& BOW + Cosine & 16.0  &  18.0  & 20.0  &  21.4 &  22.7 &  25.7 &  27.8 &  30.4 & 33.1  & 35.5  & 25.1 $\pm$ 6.2  \\ \cline{2-13}
		& BOW + SVM & 17.5  &  15.6  & 16.5  & 16.4  & 16.4  & 17.0  & 17.2  &  17.4 &  16.9 &  16.2 & 16.7 $\pm$ 0.6  \\ \cline{2-13}	
		&BOW + Softmax & 15.6  &  14.2  & 14.4  & 13.9  & 14.0  & 13.4  & 13.8  & 14.5  & 14.9  & 14.1  &  14.3 $\pm$ 0.6 \\ \cline{2-13}	
		& DBRNN-A + Softmax & 32.5   & 33.7  & 35.5  &  36.5 & 36.4  & 34.4  & 36.1  &  37.3 &  38.9 & 39.6  & \textbf{36.1  $\pm$  2.1}\\ \hline \hline

		\multirow{5}{*}{\parbox{1.4cm}{Min. train samples per class = 20}} & BOW + MNB &  21.3 &  24.3  & 30.2  & 34.8  & 38.5  &  39.4 &  37.5 &  40.7 &  42.1 &  41.8 &  35.1 $\pm$ 7.0 \\ \cline{2-13}
		
		& BOW + Cosine & 16.8  &  18.4  & 20.4  &  23.3 &  28.6 &  31.3 & 35.7  & 38.6  &  37.3 &  38.9 &  28.9 $\pm$ 8.2  \\ \cline{2-13}
		& BOW + SVM & 14.6  &  15.2  & 16.4  & 14.5  & 13.9  & 15.7  & 16.8  &  15.6 &  16.1 &  16.4 & 15.5 $\pm$  0.9 \\ \cline{2-13}
		&BOW + Softmax & 18.8  & 16.4   & 11.4  &  10.5 &  11.8 &  13.1 &  13.6 & 14.3  &  14.8 &  15.3 & 14.0 $\pm$ 2.4  \\ \cline{2-13}		
		& DBRNN-A + Softmax &  33.3  & 34.9  & 36.5  &  36.8 &  37.7 & 39.0  &  41.3 & 42.6  &  41.1 & 43.3  &  \textbf{38.8 $\pm$ 3.2} \\ \hline
	\end{tabular}
	\caption{\label{tab:deep4} Rank-$10$ accuracy obtained on the Mozilla Core project across the ten cross validations. The average accuracy over the cross validation and standard deviation is also reported. The best performing results are shown in bold.}
\end{table*}	

\begin{table*}[!h]
	\centering
	\begin{tabular}{|p{1.4cm}|l|c|c|c|c|c|c|c|c|c|c|c|}
		\hline
		\textbf{Threshold} & \textbf{Classifier}  & \textbf{CV\#1}  & \textbf{CV\#2} & \textbf{CV\#3} & \textbf{CV\#4} & \textbf{CV\#5} & \textbf{CV\#6} & \textbf{CV\#7} & \textbf{CV\#8} & \textbf{CV\#9} & \textbf{CV\#10} & \textbf{Average}\\ \hline
		\multirow{5}{*}{\parbox{1.4cm}{Min. train samples per class = 0}} & BOW + MNB & 19.1  &  21.3  & 24.5  &  22.9 & 25.8  & 28.1  & 30.3  &  31.9 &  33.94 & 35.55  & 27.4 $\pm$ 5.2  \\ \cline{2-13}
		
		& BOW + Cosine &  17.3 &  20.3  &  22.9 & 25.4  & 26.9  & 28.3  & 29.8  & 27.5  & 28.9  &  30.1 & 25.7 $\pm$ 4.1  \\ \cline{2-13}
		& BOW + SVM &  13.4 &  11.4  & 13.8  & 15.5  &  14.5 &  14.5&  14.3 &  14.4 &  14.6 & 14.6  & 14.1 $\pm$ 1.0 \\ \cline{2-13}
		&BOW + Softmax & 11.9  &  17.8  & 17.8  & 15.7  & 13.6  & 15.5  &  13.7 &  13.1 &  13.1 & 13.6  & 14.6 $\pm$ 1.9  \\ \cline{2-13}		
		& DBRNN-A + Softmax &  33.6  & 34.2  & 34.7  & 36.1  & 38.0  & 37.3  & 38.9  & 36.3  & 37.4  & 38.1  & \textbf{36.5 $\pm$ 1.7} \\ \hline \hline
		
		\multirow{5}{*}{\parbox{1.4cm}{Min. train samples per class = 5}} & BOW + MNB & 21.1  & 26.8   &  31.1 & 33.4  & 36.5  & 36.0  &  37.6 &  36.9 &  34.9 &  36.5 &  33.1 $\pm$ 5.1 \\ \cline{2-13}
		
		& BOW + Cosine & 20.8  & 23.0   & 23.7  &  26.2 &  27.4 &  29.2 &  32.3 & 32.7  & 34.1  & 35.2  & 28.5 $\pm$ 4.8 \\ \cline{2-13}
		& BOW + SVM & 14.4  &  16.0  & 17.8  &  17.8 &  17.8 & 16.3  & 16.7  & 16.7  &  16.7 &  15.2 & 16.5 $\pm$ 1.1\\ \cline{2-13}
		&BOW + Softmax &  18.2 &  14.8  & 16.7  &  16.7 & 15.4  & 14.5  & 12.5  & 12.9  &  12.9 &  13.7 & 14.8 $\pm$ 1.8 \\ \cline{2-13}		
		& DBRNN-A + Softmax &  27.6  & 34.9  &  37.9 & 38.7  & 40.1  & 42.3  & 45.2  & 44.9  &  45.0 & 44.5  & \textbf{40.1 $\pm$ 5.3} \\ \hline \hline

		\multirow{5}{*}{\parbox{1.4cm}{Min. train samples per class = 10}} & BOW + MNB &  21.7 &  27.6  & 32.1  & 34.8  &  37.7 &  34.6 &  32.6 &  34.7 & 36.7  & 38.5  &  33.1 $\pm$ 4.8\\ \cline{2-13}
		
		& BOW + Cosine & 18.1  &  21.2  & 24.4  &  27.0 &  28.3 &  30.1 &  32.3 &  34.00 & 35.4  & 36.6  & 28.7 $\pm$ 5.8\\ \cline{2-13}
		& BOW + SVM &  09.9 & 09.9   &  11.8 & 11.8  & 11.8  & 12.8  &  12.9 & 12.9  &  12.9 & 12.8  & 11.9 $\pm$ 1.1 \\ \cline{2-13}
		&BOW + Softmax & 14.3  & 15.6   & 12.1  &  09.5 &  09.5 &  11.2 & 12.0  & 12.6  &  12.1 &  12.7 & 12.1 $\pm$ 1.8 \\ \cline{2-13}	
		& DBRNN-A + Softmax & 35.1   & 36.4  & 40.5  &  42.5 & 45.4  & 47.4  & 48.9  &  49.1 & 51.1  & 51.4  & \textbf{44.8 $\pm$ 5.6} \\ \hline \hline

		\multirow{5}{*}{\parbox{1.4cm}{Min. train samples per class = 20}} & BOW + MNB & 22.0  &  22.8  & 23.6  &  26.3 & 29.2  & 32.3  &  34.4 &  36.4 &  38.6 &  38.4 & 30.4 $\pm$ 6.2 \\ \cline{2-13}
		
		& BOW + Cosine & 18.4  & 21.9 &  25.1 & 27.5  &  29.1 & 31.4  & 33.8  &  35.9 &  36.7 & 38.3  & 29.8 $\pm$ 6.3 \\ \cline{2-13}
		& BOW + SVM & 18.7  & 16.9 &  15.4 &   18.2 & 20.6  & 19.1  &  20.3 &  21.8 & 22.7  & 21.9  & 19.6  $\pm$ 2.2 \\ \cline{2-13}	
		&BOW + Softmax &  16.5 &  13.3  &  13.2 &  13.8 & 11.6  & 12.1  &  12.3 &  12.3 &  12.5 &  12.9 & 13.1 $\pm$ 1.3  \\ \cline{2-13}	
		& DBRNN-A + Softmax &  38.9  & 37.4  & 39.5  & 43.9  & 45.0  & 47.1  & 50.5  &  53.3 & 54.3  & 55.8  & \textbf{46.6 $\pm$ 6.4}  \\ \hline 
		
	\end{tabular}
	\caption{\label{tab:deep5} Rank-$10$ accuracy obtained on the Mozilla Firefox project across the ten cross validations. The average accuracy over the cross validation and standard deviation is also reported. The best performing results are shown in bold.}
\end{table*}

\subsection{Evaluation Protocol and Metric}
For a given bug report, the trained softmax classifier provides a probability value for every developer, denoting their association with the bug report. Thus, the evaluation metric that is used is the top-$k$ accuracy, which denotes the ratio of the bug reports for which the actual developer is present in the top-$k$ retrieved results. Across the cross validation (CV) sets, varying classes or a set of developers are used. Thus during CV\#1, the classes used for training and testing is different from the classes used in CV\#2. Thus, as the classifier model across the CV is trained on different classes, taking the average accuracy would only provide a ballpark number of the performance, while is not accurately interpretable. Thus, it is required to report the top-$k$ accuracy of each cross validation set to understand the variance introduced in the model training~\cite{kohavi1995study}.

For learning the deep representation, a DBRNN-A is constructed having $300$ LSTM units and the dropout probability is $0.3$. A categorical cross entropy based loss function is used with Adam optimizer, learning rate as $0.001$, and trained for $100$ epochs with early stopping. The model architecture and parameters utilized are shown in Fig.~\ref{fig:dbrnna}

\subsection{Comparison with Existing Algorithms}
The major challenge in cross comparison of algorithm performance is the lack of a public benchmark dataset and source code implementation of the existing research. Thus, the bug triaging accuracy obtained in the previous research works cannot be compared with the proposed approach, unless the results are shown in the same dataset.
% write about using only table 4 data
Thus, we implement some of the successful approaches for automated bug triaging from the literature~\cite{anvik2011reducing}~\cite{xuan2015towards}~\cite{Jonsson2016} and compare it with our algorithm using our benchmark dataset. Term frequency based BOW is used to represent the combined title and description from a bug report, as shown in Fig.~\ref{fig:rnn}.
Using these features, we evaluate the performance of four different classifiers: (i) Softmax classifier~\cite{schmidt2013minimizing}, (ii) Support Vector Machine (SVM)~\cite{wu2004probability}, (iii) Multinomial Naive Bayes (MNB)~\cite{kibriya2004multinomial}, and (iv) Cosine distance based matching~\cite{mihalcea2006corpus}. The four supervised classifiers are implemented using the Python \emph{scikit-learn}\footnote{http://scikit-learn.org/} package. All these four classifiers use only the classifier training and testing data and do not use the untriaged bug reports. 

\begin{figure}[ht]
	\centering
	\includegraphics[width=3.3in]{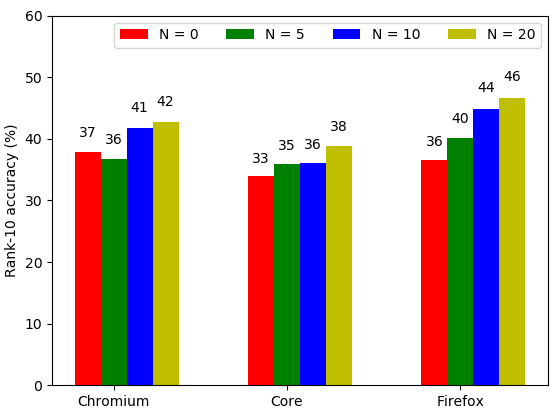}
	\caption{The rank-$10$ average accuracy of the deep learning algorithm on all three datasets. It can be observed that as the number of training samples per class increases, the overall triaging accuracy increases, addressing \textbf{RQ3}.}
	\label{fig:viz1}
\end{figure}

\begin{figure}[ht]
	\centering
	\includegraphics[width=3.2in]{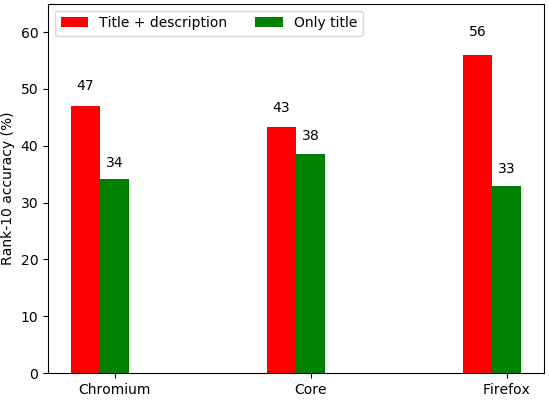}
	\caption{The rank-$10$ average accuracy of the deep learning algorithm on all three datasets by using only title or title along with the description in bug report. Discarding the description reduces the performance significantly, addressing \textbf{RQ4}.}
	\label{fig:viz2}
\end{figure}

\subsection{Result Analysis}
The results obtained in the Google Chromium, Mozilla Core, and Mozilla Firefox datasets are shown in Table~\ref{tab:deep3}, Table~\ref{tab:deep4}, and Table~\ref{tab:deep5}, respectively. The main research questions focused in this paper are answered using the obtained results.

\noindent\textbf{RQ1: Is it feasible to automated perform bug triaging using deep learning?} \\
From the obtained results, it can be clearly observed that the deep learning the representation of a bug report is a feasible and potentially competent approach for bug triaging. The proposed DBRNN-A approach provided rank-10 triaging accuracy in the range of $34 - 47\%$. 
All the experiments are executed in an Intel(R) Xeon(R) CPU E5-2660 v3, running at 2.60GHz and with a Tesla K80 GPU. Learning the feature representation model and training the classifier are usually offline tasks and do not contribute towards the testing time. For example in the Google Chroimum dataset, training the DBRNN-A takes about 300 seconds per epoch. For the entire CV\#10 subset, training and testing time the softmax classifier takes about 121 seconds and 73 seconds, respectively. However, after training the models, developer assignment for a new bug report takes only 8 milliseconds using the proposed approach (feature extraction + classification), highlighting the speed of the proposed approach. 

\noindent\textbf{RQ2: How does the unsupervised feature engineering approach perform, compared to traditional feature engineering approaches?} \\
It can be concretely observed from the results that the feature learning using DBRNN-A outperforms the traditional BOW feature model. In Chromium dataset, rank-$10$ average accuracy of BOW + Softmax is around $9-12\%$, while the best performing classifier provides $26-28\%$. This shows the challenging nature of the bug triaging problem in the large dataset that we have created. However, DBRNN-A provides a rank-$10$ average accuracy in the range of $37-43\%$ improving results by $12-15\%$. Similarly in Mozilla Core, we observe a $3-5\%$ improvement and in Mozilla Firefox, we observe a $7-17\%$ improvement in rank-$10$ average accuracy by using deep learning features because the deep learning model could memorize the context from longer sentences in a bug report reasoning for the large improvement in performance. From the results, we observe that for BOW features MNB and cosine distance based matching outperforms SVM and softmax classifier. Although SVM is a popular choice of a supervised classifier, for real numbered sparse features in BOW model, feature independence which is assumed both in MNB and cosine distance matching proves successful.

\noindent\textbf{RQ3: Is there an effect on the number of training samples per class on the performance of the classifier?} \\
Both intuitively and experimentally, we find that as the minimum number of training samples per class increased, the performance of the classification improved across all the bug repositories by learning better classification boundary. For instance in Chromium dataset, when a classifier is trained with threshold as zero, DBRNN-A produced an average rank-$10$ accuracy of $37.9\%$ and steadily increased to $42.7\%$ when threshold is $20$. Fig~\ref{fig:viz1} captures the improvement in rank-$10$ average accuracy for all the three dataset. However, for the collected data having a threshold greater than $20$ did not improve the classification accuracy. Also, as we proceed from CV\#1 from CV\#10, we observe that the performance of DB-RNN increases. Despite the fact that there is increased number of testing classes, the availability of increased training data improves the classification performance. Thus, empirically the more training data is available for the classifier, the better is the performance. Also, across the cross validations there is about $(2-7)\%$ standard deviation in all dataset. This emphasizes the importance of studying the performance of each cross validation set along with the average accuracy.

\noindent\textbf{RQ4: What is the effect of using only the title (or summary) of the bug report in performing triaging compared with the using the description as well ?}\\
The performance of the deep learning model was studied by using only the title (summary) of the bug report and completely ignoring the description information. The experiments were conducted on all three datasets, with the minimum number of train samples N=$20$ and CV\#10. Fig.~\ref{fig:viz2} compares the rank-$10$ average accuracy on all three datasets with and without using the description content. It can be clearly observed that discarding description significantly reduces the performance of triaging of up to $23\%$ in Firefox dataset.

\noindent\textbf{RQ5: Is transfer or cross-data learning effective using deep learning, where the deep learning model is trained using one dataset and used to perform triaging in another dataset?} \\
Transfer learning reduces the offline training time significantly by re-using a model trained using another dataset. However, most of the models fail while transferring the learnt model across datasets. The effectiveness of the deep learning model in transfer learning is studied, by training the model in Chromium dataset and testing in Core and Firefox datasets (Mozilla dataset). Using the deep learning model trained on Chromium dataset, the average rank-$10$ accuracy obtained when N=$20$ are $42.7\%$ for Chromium test set, $39.6\%$ for Core test set, and $43\%$ on Firefox test set. The obtained results are comparable with the results obtained by training and testing on the same dataset. This shows that the proposed approach is capable of using a model trained on dataset to triage bug reports in another dataset, effectively. 

\section{Threats to Validity}
 There are certain threats to establish the validity of the proposed results. While the common threats to any learning based classification system are applicable, few of these threats are specific to the advocated problem, as follows:
\begin{enumerate}
	\item The results are shown using three open source bug repositories with different characteristics to ensure generalization. However, commercial bug tracking systems may follow different patterns and hence our results may not be directly extensible to such repositories.
	\item Currently in our approach, only the bug report title and description are considered. While the experimental results show that these two unstructured text data are necessary, there could be other added information required to sufficiently triage a bug report.
	\item For comparison purpose, we re-implemented some of the successful algorithms in the literature for bug triaging, as there is no publicly implementation available. Although we have implemented the algorithms true to the best of our understanding, there could be some minor deviations from the original implementation.
	\item Both during training and testing of a classifier, we assumed only one developer as the rightful owner of a bug report. However, based on patterns and history of bug reports that are solved, there could be more than one active developer in the project who could potentially address the bug. 
\end{enumerate}

\begin{table*}[ht]
	\centering
	\begin{tabular}{|p{2.1cm}|p{3.5cm}|p{1.7cm}|p{2.3cm}|p{2.6cm}|p{3.2cm}|}
		\hline
		\textbf{Paper} & \textbf{Information used} & \textbf{Feature extracted} & \textbf{Approach} & \textbf{Dataset} & \textbf{Performance} \\ \hline
		
		Bhattacharya et al., 2010 \cite{bhattacharya2010fine} & title, description, keywords, product, component, last developer activity & tf-idf + bag-of-words & Naive Bayes + Tossing graph & Eclipse\# 306,297  \newline  Mozilla\# 549,962 &  Rank\#5 accuracy 77.43\%  \newline  Rank\#5 accuracy 77.87\% \\ \hline 
		
		Tamrawi et al., 2011 \cite{tamrawi2011fuzzy} & title, description & terms & A fuzzy-set feature for each word & Eclipse\# 69829 & Rank\#5 accuracy 68.00\% \\ \hline
		
		Anvik et. Al., 2011 \cite{anvik2011reducing} & title, description & normalized tf & Naive Bayes, EM, SVM, C4.5, nearest neighbor, conjunctive rules & Eclipse\# 7,233  \newline \newline Firefox\# 7,596 & Rank\#3 prec. 60\%, recall 3\% \newline Rank\#3 prec. 51\%, recall 24\% \\ \hline
		
		Xuan et. Al., 2012 \cite{xuan2012developer}	& title, description	& tf-idf, developer prioritization & Naive Bayes, SVM & Eclipse\# 49,762  \newline \newline Mozilla\# 30,609 & Rank\#5 accuracy 53.10\% \newline \newline Rank\#5 accuracy 56.98\%  \\ \hline
		
		Shokripour et al. 2013 \cite{shokripour2013so} &  title, description, detailed source code info & weighted unigram noun terms & Bug location prediction + developer expertise & JDT-Debug\# 85  \newline \newline Firefox\# 80 & Rank\#5 accuracy 89.41\%  \newline \newline Rank\#5 accuracy 59.76\%
		\\ \hline
		
		Wang et al., 2014 \cite{wang2014fixercache} & title, description & tf & Active developer cache & Eclipse\# 17,937 \newline  Mozilla\# 69,195 & Rank\#5 accuracy 84.45\% \newline  Rank\#5 accuracy 55.56\% \\ \hline
		
		Xuan et. al., 2015 \cite{xuan2015towards} & title, description & tf & feature selection with Naive Bayes & Eclipse\# ~50,000  \newline  Mozilla\# ~75,000 & Rank\#5 accuracy 60.40\%   \newline Rank\#5 accuracy 46.46\% \\ \hline		
		
		Badashian et. al., 2015 \cite{badashian2015crowdsourced} & title, description, keyword, project language, tags from stackoverflow, github & Keywords from bug and tags & Social expertise with matched keywords & 20 GitHub projects, 7144 bug reports & Rank\#5 accuracy 89.43\% \\ \hline
		
		Jonsson et. al., 2016~\cite{Jonsson2016} & title, description & tf-idf & Stacked Generalization of a classifier ensemble & Industry\# 35,266 & Rank\#1 accuracy 89\% \\ \hline
		
	\end{tabular}
	\caption{\label{tab:tablex} Summary of various machine learning based bug triaging approaches available in literature, explaining the features and approach used along with its experimental performance.}
\end{table*}

\section{Other Applications}
The bug representation is learnt directly from the data in an unsupervised fashion and the features are task independent. This gives us a flexibility to use these features to learn a supervised classifier for any task or application. We have discussed a few of other possible applications for the proposed feature representation.

\begin{itemize}
	\item Which bug gets fixed: It is a challenging research problem that have been addressed in literature~\cite{guo2010characterizing}. Using the same feature representation extracted in this research, a supervised or semi-supervised binary classifier can be trained to classify fixed bugs with non-fixed bugs.
	\item Bug-fix time prediction: As Bhattacharya et al.~\cite{bhattacharya2011bug} discuss, a predictive model can be constructed using the proposed features to learn the time required to fix the bug.
	\item Reopen analysis: It provides an interesting insight from the maintenance perspective to study which bugs get reopened during its lifecycle. As discussed by Zimmermann et al.~\cite{zimmermann2012characterizing}, characterizing these bugs and predicting them can be performed using the deep learning features.
	\item Bug priority estimation: Priority of the bug is to be estimated before triaging happens~\cite{bhattacharya2011bug}. Based on the priority, the SLA clock for the bug and the developer to be assigned might change. A 5-point scale priority can be formulated as a five class classification and using the learnt features, a supervised classifier can be learnt.
\end{itemize}

\section{Related Work}
Table \ref{tab:tablex} presents a list of closely related works on bug triaging arranged in a chronological order (year $2010$ to $2016$). A majority of previous techniques have used bug summary/title and description \cite{anvik2011reducing} \cite{tamrawi2011fuzzy} \cite{xuan2015towards} \cite{xuan2012developer} because they are available at the time of ticket submission and do not change in tickets' lifecyle. Bhattacharya et. al. \cite{bhattacharya2010fine} use additional attributes such as product, component, and the last developer activity to shortlist developers. Shokripour et al. \cite{shokripour2013so} use code information for improved perfomance. Badashian et. al. \cite{badashian2015crowdsourced} identify developers' expertise using stack overflow and keywords from bug description. 

From table \ref{tab:tablex}, we observe that many different feature models such as tf, normalized tf, tf-idf, and n-grams have been employed. Choosing which feature model to use is an engineering design choice and it is challenging to choose which feature model will best represent the collected data. In this research, we address this challenge and design a deep bidirectional RNN which directly learns the best feature representation from the data in an unsupervised fashion. Further, we move beyond a word level representation model and propose a sequence of word representation model, to learn a unified representation for the entire bug report.

\section{Conclusion}
In this research we proposed a novel software bug report (title + description) representation algorithm using deep bidirectional Recurrent Neural Network with attention (DBRNN-A). The proposed deep learning algorithm learns a paragraph level representation preserving the ordering of words over a longer context and also the semantic relationship. The performance of four different classifiers, multinomial naive Bayes, cosine distance, support vector machines, and softmax classifier are compared. To perform experimental analysis, bug reports from three popular open source bug repositories are collected - Google Chromium (383,104), Mozilla Core (314,388), and Mozilla Firefox (162,307). Experimental results shows DBRNN-A along with softmax classifier outperforms the bag-of-words model, improving the rank-$10$ average accuracy in all three datasets. Further, it was studied that using only the title information for triaging significantly reduces the classification performance highlighting the importance of description. The transfer learning ability of the deep learning model is experimentally shown, where the model learnt on the Chromium dataset competitively triaged the bugs in the Mozilla dataset. Additionally, the dataset along with its complete benchmarking protocol and the implemented source code is made publicly available to increase the reproducibility of this research.

\bibliographystyle{plain}
\bibliography{bug_triage}{}

\end{document}